\documentclass{ws-ijmpd}

\begin{document}

\def\nocropmarks{\vskip5pt\phantom{cropmarks}}

\markboth{S. G. Ghosh and D. W. Deshkar} {Gravitational Collapse
of perfect fluid in higher dimensional space-times}

\catchline{}{}{}

\title{GRAVITATIONAL COLLAPSE OF PERFECT
FLUID IN SELF-SIMILAR HIGHER DIMENSIONAL SPACE-TIMES}

\author{\footnotesize S. G.~GHOSH $^1$
\footnote{Author to whom all correspondence should be directed.}
~and ~D. W.~DESHKAR}

\address{Centre for Mathematical Physics,  Science College \\
Congress Nagar  Nagpur-440 012, India\\ $^1$ sgghosh@iucaa.ernet.in}%

\maketitle

\pub{Received (received date)}{Revised (revised date)}

\begin{abstract}
We investigate the occurrence and nature of naked singularities in
the gravitational collapse of an adiabatic perfect fluid
in self-similar higher dimensional space-times. It is shown that strong
curvature naked singularities could occur if the weak energy condition holds.
Its implication for cosmic censorship conjecture is discussed.  Known results
of analogous studies in four dimensions can be recovered.
\end{abstract}


\section{Introduction}
The cosmic censorship conjecture, which is widely believed to be
true, put forward by Penrose \cite{rp} thirty years ago has become
the corner stone of general relativity.  Moreover, it is being
envisaged as a basic principle of nature.  The conjecture (in its
weak form) asserts that  no physically reasonable matter field can
give rise to singularities which are observable from asymptotic
regions of the space-time. According to the strong version of the
conjecture, such singularities are invisible to all observers,
i.e., there occurs no naked singularities for any observer, even
someone close to it.

However, despite the flurry of activity over the years, the
validity of this conjecture is still an open question and very far
from being settled. In view of this, it is certainly worthwhile
exercise to search for examples which appear to violate the
conjecture (for recent reviews and reference on conjecture, see
\cite{psj,r2}).  There are of course number of solutions of
Einstein's equations, in ,
  which admit naked singularities (see, e.g., \cite{psj} and
reference therein). This happens in the gravitational collapse of
different geometrical configuration and various types of matter.
Those received most attention, and which can arise from regular
initial data, are generated in self-similar collapse
\cite{op,op1,wl,hp,dj,jl,bn}. The self-similar models may be refer
to highly symmetric and specialized situation.  But the
self-similarity is interesting to study as it reduces Einstein's
field equations to a system of ordinary differential equations and
it
 may simplify physical interpretation of models.

In recent years, superstring and field theories provoked great
interest among theoretical physicists in studying physics of
higher dimensions \cite{rs} and hence significant efforts have
been expended to study gravitational collapse models in higher
dimensions (HD) \cite{hd} but only a few, for example see
\cite{gd}, have studied these problems from viewpoint of the
cosmic censorship conjecture. It is  shown that a naked
singularity forms in the HD spherical collapse of null radiation
\cite{gd} and of dust as well \cite{gb}. However, these models
neglect pressure which is likely to play important role in final
outcome of the collapse.   Ori and Piran \cite{op}, in 4D case,
have studied numerically self-similar collapse of an adiabatic
perfect fluid and have shown that with a soft equation of state
the collapse can give rise to a naked singularity \cite{op1}. This
was followed by analytical discussions of spherical collapse of
perfect fluid under assumption of self-similarity \cite{dj}. It is
therefore interesting to study gravitational collapse of perfect
fluid in HD space-times. There is some literature on HD perfect
fluid collapse \cite{gss} from the viewpoint of the conjecture.
However, these studies are restricted to 5D space-time and cannot
be reduced to the 4D case.

 The main purpose of this paper is to  generalize analytical
studies,  of self-similar collapse of adiabatic perfect fluid
collapse,  in 4D to $D = n+2$-dimensional space-times (where
$n\geq2$) and to show  that previous results of
 gravitational collapse, obtained in 4D space-time,
are also valid in HD space-time.  Infact, It is shown that
gravitational collapse of HD space-times gives rise to shell
focusing globally naked strong curvature singularities.  This is
done in section IV. In section II we write the effective
Einstein's equations in HD and generalize relevant Cahill and Taub
\cite{ct} solutions. We have also done matching of these solutions
to the HD Vaidya solutions in sections III and, we conclude with a
discussion.

\section{HD perfect fluid space-times}
To facilitate the discussion and set notation, we start with HD
self-similar spherically symmetric space-times, which  in comoving
coordinates reads:
\begin{equation}
ds^2 = e^{ \vartheta}dt^2 - e^{ \chi}dr^2 - r^2 S^2 d \Omega^2
\label{eq:me}
\end{equation}
where $\vartheta$, $\chi$ and $S$ are function of $r$ and $t$ and
\begin{eqnarray}
d\Omega^2& = & \sum_{i=1}^{n} \left[ \prod_{j=1}^{i-1} \sin^2
\theta_j \right] d \theta_i^2 =
  d \theta_1^2+  \sin^2 \theta_1 d \theta_2^2
+ \sin^2 \theta_1 \sin^2 \theta_2 d\theta_3^2+\;.\; .\; .\;  +
\nonumber \\
& &
 \sin^2 \theta_1 \sin^2 \theta_2 .\; .\; . \sin^2
\theta_{n-1} d\theta_n^2 \label{eq:ns}
\end{eqnarray}
is the metric on an $n$-sphere and $n = D-2$, where $D$ is the
total number of dimensions.  For a perfect fluid, the Einstein
equations are
\begin{equation}
R^{ab} - \frac{1}{2} R g^{ab} = (\rho+P)u^{a}u^{b} - P g^{ab}
\label{ee}
\end{equation}
where $\mu$ is energy density, $P$ is pressure,  $u_a = e^{-
\vartheta/2}\delta_a^t$ is comoving $n$-velocity, and we choose
units with $8 \pi G = c =1$. A spherically symmetric self-similar
solution is one in which the  space-time admits a homothetic
Killing vector $\xi$ which satisfies

\begin{equation}
\mathcal{L}_{\xi}g_{ab} =\xi_{a;b}+\xi_{b;a} = 2 g_{ab}
\end{equation}
where $\mathcal{L}$ denotes the Lie derivative.   This means that
by a suitable transformation of coordinates, all metric
coefficients and dependent variables can be put in the form in
which they are functions of a single independent variable (say z),
which is dimensionless combination of space and time coordinates
and hence $\vartheta$, $\chi$ and $S$ are functions of the
self-similarity variable $z=t/r$.

The simplest equation of state compatible with similarity solution
is one of the form $p(z) = \alpha \zeta(z)$, where $0 \leq \alpha
\leq 1$ is a constant.

The conservation equations $T^{ab}_{;b}=0$
 are immediately integrated to give
\begin{equation}
e^{\vartheta} = \gamma (\zeta z^2)^{-2 \alpha/(1+\alpha)}
\label{en}
\end{equation}
\begin{equation}
e^{\chi} =\eta (\zeta )^{-2 /(1+\alpha)}S^{-2 n} \label{ep}
\end{equation}
where $\gamma$ and $\eta$ are integration constants.  The
remaining field equations reduce to set of ordinary differential
equations:

\begin{eqnarray}
G^t_t & = & - e^{-\vartheta} \left[\frac{n(n-1)}{2}
\frac{\dot{S}^2}{S^2} + \frac{n}{2} \frac{\dot{S}}{S}
\dot{\chi}\right] + e^{-\chi}\Big[\frac{n(n-1)}{2}  +
n\frac{\ddot{S}}{S} z^2 + \frac{n(n-1)}{2}
\frac{\dot{S}^2}{S^2}z^2 - n(n-1) \frac{\dot{S}}{S}z + \nonumber \\
& & \frac{n}{2} \dot{\chi} z  - \frac{n}{2} \frac{\dot{S}}{S}
\dot{\chi} z^2 \Big]-\frac{n(n-1)}{2} \frac{1}{S^2} = -\zeta
\label{eq:gtt}
\end{eqnarray}
\begin{eqnarray}
G^r_r & = &  -  e^{-\vartheta} \left[  n\frac{\ddot{S}}{S} -
\frac{n}{2} \frac{\dot{S}}{S} \dot{\vartheta} + \frac{n(n-1)}{2}
\frac{\dot{S}^2}{S^2} \right] + e^{-\chi}  \Big[\frac{n(n-1)}{2} +
\frac{n(n-1)}{2}
\frac{\dot{S}^2}{S^2}z^2 - n(n-1) \frac{\dot{S}}{S}z  \nonumber\\
&&+\frac{n}{2} \dot{\vartheta} z + \frac{n}{2} \frac{\dot{S}}{S}
\dot{\vartheta} z^2 \Big]-\frac{n(n-1)}{2} \frac{1}{S^2} =
p\label{eq:grr}
\end{eqnarray}

\begin{eqnarray}
G^{\theta_1}_{ \theta_1} & = & e^{-\vartheta}
\left[\frac{\ddot{\chi}}{2} + \frac{\dot{\chi}^2}{4} + (n-1)
\frac{\ddot{S}}{S} -
 (n-1) (n-2) \frac{\dot{S}^2}{S^2} +  \frac{(n-1)}{2} (\dot{\chi} -
\dot{\vartheta})  \frac{\dot{S}}{S} - \frac{\dot{\chi}
\dot{\vartheta}}{4} \right]
\nonumber \\
&& + \; e^{-\chi} \Big[ - \frac{\ddot{\vartheta} z^2 }{2} +
\dot{\vartheta} z + \frac{\ddot{\vartheta}^2 z^2 }{4} + (n-1)
\frac{\ddot{S} z^2}{S} - \frac{(n-1)(n-2)}{2} \frac{\ddot{S}^2 \;
z^2}{S^2} - (n-1)(n-2) \frac{\ddot{S} \; z}{S} +
\nonumber \\
&& \frac{(n-1)}{2} (\dot{\chi} -  \dot{\vartheta}) \left(1 -
\frac{\ddot{S} \; z}{S} \right) z - \dot{\chi} \dot{\vartheta} z^2
+ \frac{(n-1)(n-2)}{2} \Big] - \frac{(n-1)(n-2)}{2} \frac{1}{S^2}
= p \label{eq:gth}
\end{eqnarray}

\begin{equation}
G^{\theta_1}_{\theta_1} = G^{\theta_2}_{\theta_2} = \; . \; . \; . \; =
G^{\theta_n}_{\theta_n}
\end{equation}
\begin{equation}
G^t_r =\frac{n}{2}\frac{z}{S} \left[2
\ddot{S}-\dot{S}\dot{\vartheta}- \dot{S}\dot{\chi}+\frac{S
\dot{\chi}}{z} \right]=0
 \label{eq:gtr}
\end{equation}
where the overdot denote the derivative with respect similarity
parameter $z$. We have assumed the following expressions for
pressure and energy
\begin{equation}
P = \frac{p(z)}{r^2}, \hspace{.5in} \rho =
\frac{\zeta(z)}{r^2}\label{edp}
\end{equation}
Elimination of $ \ddot{S}$ from Eqs. (\ref{eq:gtt}) and
(\ref{eq:grr}) leads to
\begin{eqnarray}
& & \frac{n(n-1)}{2} \left( \frac{\dot{S}}{S}\right)^2 V
+\frac{n}{2} \frac{\dot{S}}{S}\left(\dot{V} + 2 n z e^{\vartheta}
\right) + e^{ \chi + \vartheta}  \left[-{\zeta}- \frac{n(n-1)}{2}
e^{- \chi}+ \frac{n(n-1)}{2} \frac{1}{S^2} \right] \nonumber \\
&&  = 0 \label{sd}
\end{eqnarray}
and
\begin{equation}
\dot{V(z)} = \frac{2}{n}z e^{\vartheta} \left[ (\zeta+p) e^{\chi}
-n \right] = \frac{2}{n}z e^{\vartheta} (\Theta-n) \label{vd}
\end{equation}
where the quantities $V$ and $\Theta$ are defined as
\begin{equation}
V(z) = e^{\chi} - z^2 e^{\vartheta}, \hspace{.3in} \Theta(z) =
(\zeta+p)e^{\chi}
\end{equation}
One can also write,  $ \Theta =  r^2 e^{\chi} \left(T^0_0 - T^1_1
\right)$. Thus the matter satisfy weak energy condition: $\rho
\geq 0$ and $\rho + p \geq 0 $ \cite{he} if and only if $T_{ab}
W^a W^b \geq 0 $, for all nonspacelike vector $W^a$, which implies
that $\Theta(z) \geq 0$ for all $z$.

\section{Junction conditions}
We consider a spherical surface with its motion described by a
time-like $n+1$ surface $\Sigma$, which divides space-times into
interior and exterior manifolds.  We shall first cut the
space-times along time like hypersurface, and then join the
internal part with the outgoing HD
 Vaidya solutions.  The metric on
the whole space-times can be written in the form
\begin{equation}
ds^2 = \left\{ \begin{array}{ll}
         e^{\vartheta}dt^2 -  e^{\chi}dr^2 - r^2 S^2 d \Omega^2, & \mbox{$r \leq r_{\Sigma}$}, \\
        \left[ 1 -  \frac{2 m(v)}{(n-1){\bf r}^{(n-1)}} \right] dv^2 + 2 dv d{\bf r} - {\bf r}^2 d \Omega^2, & \mbox{$r \geq  r_{\Sigma}$}.
                \end{array}
        \right.                         \label{eq:mv}
\end{equation}
The metric on the hypersurface $r=r_{\Sigma}$ is given by
\begin{equation}
ds^2 =  d\tau^2 - {\mathcal{R}}^2(\tau) d\Omega^2 \label{bm}
\end{equation}
For the junction conditions, we suitably modify the approach given
in \cite{no,rw} for our HD case.  Hence we have to demand
\begin{equation}
(ds_{-}^2)_{\Sigma} = (ds_{+}^2)_{\Sigma} = (ds^2)_{\Sigma}
\label{jc1}
\end{equation}
The second junction condition is obtained by requiring the
continuity of the extrinsic curvature of $\Sigma$ across the
boundary.  This yields
\begin{equation}
K^-_{ij} = K^+_{ij} \label{jc2}
\end{equation}
where $K^{\pm}_{ij}$ is extrinsic curvature to $\Sigma$, given by
\begin{equation}
K^{\pm}_{ij} = - n_{\alpha}^{\pm} \frac{\partial^2
x^{\alpha}_{\pm}}{\partial \xi^i \partial \xi^j} -
n_{\alpha}^{\pm} \Gamma^{\alpha}_{\beta \gamma} \frac{\partial
x^{\beta}_{\pm} }{\partial \xi^i} \frac{\partial x^{\gamma}_{\pm}
}{\partial \xi^j}
 \label{ec}
\end{equation}
and where $\Gamma^{\alpha}_{\beta \gamma}$ are Christoffel
symbols, $n^{\pm}_{\alpha}$ the unit normal vectors to $ \Sigma $,
$x^{\alpha}$ are the coordinates of the interior and exterior
space-time and $\xi^i$ are the coordinates that defines $\Sigma$.
From the junction condition (\ref{jc1}), we obtain
\begin{equation}
\frac{dt}{d\tau} = \frac{1}{e^{\vartheta(r_{\Sigma},t)/2}}
\label{jc3}
\end{equation}
\begin{equation}
r_{\Sigma} S(r_{\Sigma}, t) = {\bf r}(\tau) \label{jc4}
\end{equation}
\begin{equation}
\left( \frac{dv}{d\tau} \right)^{-2}_{\Sigma} = \left[1 -
 \frac{2 m(v)}{(n-1){\bf r}^{(n-1)}}  + 2 \frac{d{\bf r}}{dv} \right]
 \label{jc5}
\end{equation}
The non-vanishing components of intrinsic curvature $K_{ij}$ of
$\Sigma$ can be calculated and the result is

\begin{eqnarray}
&& K^{-}_{\tau \tau} = \left(- e^{- \chi/2} \vartheta_r
\right)_{\Sigma}
\label{ecia} \\
&& K^{-}_{\theta _1\theta_1}  = \left[ e^{- \chi /2} r S
\left(S + r S_r \right) \right]_{\Sigma}  \label{ecib} \\
&& K^{+}_{\tau \tau} = \left[ \frac{d^2v}{d\tau^2}
\left(\frac{dv}{d\tau} \right)^{-1} - \left(\frac{dv}{d\tau}
\right) \frac{m(v)}{{\bf r}^n}   \right]_{\Sigma}  \label{ecic}\\
&& K^{+}_{\theta_1 \theta_1} = \left[ {\bf r} \frac{d{\bf r}}{d \tau}
 + \left(\frac{dv}{d\tau} \right)
\left(1 - \frac{2 m(v)}{(n-1){\bf r}^{(n-1)}}  \right) {\bf r} \right]_{\Sigma} \label{ecid}  \\
&& K^{\pm}_{\theta_i \; \theta_i} = \sin^2 \theta
K^{\pm}_{\theta_{i-1}\; \theta_{i-1}} \label{ecie}
\end{eqnarray}

where subscripts $r$ and $t$ denote partial derivative with
respect to $r$ and $t$ respectively.  The unit normal to the
$\Sigma$ are given by
\begin{equation}
n_{\alpha}^- = (0, e^{\chi(r_{\Sigma},t)/2}, 0, . . ., 0)
\label{unm}
\end{equation}
\begin{equation}
n_{\alpha}^+ = \left[1 -\frac{2 m(v)}{(n-1){\bf r}^{(n-1)}}   +
2  \frac{d {\bf
r}}{dv} \right]^{-1/2} \left(- \frac{d {\bf r}}{dv}, 1, 0, . . ., 0
\right) \label{unp}
\end{equation}
From Eqs. (\ref{jc2}), (\ref{ecib}) and (\ref{ecid}) we have
\begin{equation}
\left[ \left(\frac{dv}{d\tau} \right) \left(1 - \frac{2
m(v)}{(n-1){\bf r}^{(n-1)}} \right) {\bf r} + {\bf r}\frac{d{\bf
r}}{d\tau} \right] = \left[ e^{- \chi/2} r S (S + r S_r)
\right]_{\sum} \label{mf1}
\end{equation}
With the help of Eqs. (\ref{jc3}), (\ref{jc4}) and (\ref{jc5}), we
can write Eq. (\ref{mf1}) as
\begin{equation}
m (v) = \frac{n-1}{2}(rS)^{(n-1)} \left[1 + \frac{r^2
S_t^2}{e^{\vartheta}} - \frac{(S + r S_r)^2}{e^{\chi}}
\right]\label{mf}
\end{equation}
which is the total energy entrapped inside the surface $\Sigma$.
This expression  (\ref{mf}) is HD generalization of well known
mass function introduced by Cahill and McVittie \cite{cm}. From
Eqs. (\ref{ecia}) and  (\ref{ecic}), using  (\ref{jc3}), we have
\begin{equation}
\left[ \frac{d^2v}{d\tau^2} \left(\frac{dv}{d\tau} \right)^{-1} -
\left(\frac{dv}{d\tau} \right) \frac{m(v)}{{\bf r}^n}
\right]_{\Sigma} = - \left( \frac{\vartheta_r}{2 e^{\chi/2}}
\right)_{\Sigma} \label{jc}
\end{equation}
Substituting Eqs. (\ref{jc3}), (\ref{jc4}) and (\ref{mf}) into
(\ref{mf1}) we can write
\begin{equation}
\left(\frac{dv}{d\tau} \right)_{\Sigma}= \left[ \frac{S + r
S_r}{e^{\chi/2}} + \frac{r S_t}{e^{\vartheta/2}}
\right]^{-1}_{\Sigma} \label{vd1}
\end{equation}
Differentiating (\ref{vd1}) with respect to $\tau$ and using Eqs.
(\ref{mf}), we can rewrite (\ref{jc}) as
\begin{eqnarray}
& & - \left(\frac{\vartheta_r}{2 e^{\chi/2}}  \right)_{\Sigma}=
\Big[ \Big[ - \frac{r}{e^{\chi/2}} S_{tr} + \frac{\chi_t (S + r
S_r)}{2 e^{\chi/2}} + \frac{r \vartheta_t S_t}{2 e^{\vartheta/2}}
- \frac{r S_{tt}}{e^{\vartheta/2}} - \frac{n-1}{2} \frac{r
S_{t}^2}{e^{\vartheta/2} S} - \nonumber \\
& & \frac{S_t}{e^{\chi}} + \frac{(n-1)e^{\vartheta/2}}{rS} \times
\left(\frac{(S + r S_r)^2}{e^{ \chi}} -1 \right) \Big] \left[
\frac{S + r S_r}{e^{\chi/2}} + \frac{r S_t}{e^{\vartheta/2}}
\right]^{-1} \frac{1}{e^{\vartheta/2}} \Big]_{\Sigma} \label{jc7}
\end{eqnarray}
Next we translate the above equation in terms of $z = t/r$, which
is given by
\begin{eqnarray}
&&  -  e^{-\vartheta} \left[  n\frac{\ddot{S}}{S} - \frac{n}{2}
\frac{\dot{S}}{S} \dot{\vartheta} + \frac{n(n-1)}{2}
\frac{\dot{S}^2}{S^2} \right] + e^{-\chi}\Big[\frac{n(n-1)}{2} +
\frac{n(n-1)}{2} \frac{\dot{S}^2}{S^2}z^2 - n(n-1)
\frac{\dot{S}}{S}z + \nonumber \\
&& \frac{n}{2} \dot{\vartheta} z + \frac{n}{2} \frac{\dot{S}}{S}
\dot{\vartheta} z^2 \Big]- \frac{n(n-1)}{2} \frac{1}{S^2} = -
\frac{n}{2}\frac{z e^{-(\chi +\vartheta)/2}}{S} \left[2
\ddot{S}-\dot{S}\dot{\vartheta}-
\dot{S}\dot{\chi}+\frac{S \dot{\chi}}{z} \right] \nonumber \\
&& \label{eq:grr1}
\end{eqnarray}
Comparing (\ref{eq:grr1}) with (\ref{eq:grr}) and (\ref{eq:gtr}),
we can finally write
\begin{equation}
(P)_{\Sigma}=0 \label{pz}
\end{equation}
Eq. (\ref{pz}) shows  that
the pressure will vanish at the boundary which implies
 radiation cannot exist and exterior space-time  $\mathcal{V}_E$ is
HD Schwarzschild space-time.

\section{The structure of singularities in HD spherical collapse}
  Let $K^{a} = dx^a/dk$ be the
tangent vector to the null geodesics, where $k$ is an affine
parameter. Then $g_{ab}K^a K^b=0$ and it follows that along null
geodesics, we have $ \xi^a K_{a} = C$, where $C$ is a constant.
From this algebraic equation and the null condition, we get the
following exact expressions for $K^t$ and $K^r$:
\begin{equation}
K^t = \frac{C \left[z \pm e^{\chi} \Pi\right]}{r\left[ e^{ \chi} -
e^{\vartheta} z^2 \right]} \label{kt1}
\end{equation}
\begin{equation}
K^r = \frac{C \left[1 \pm z e^{\vartheta} \Pi\right]}{r\left[ e^{
\chi} - e^{\vartheta} z^2 \right]} \label{kr1}
\end{equation}
where $\Pi =\sqrt{e^{-\chi - \vartheta}} > 0 $. We shall follow
mainly arguments of Joshi and Dwivedi \cite{dj} in our subsequent
analysis.  Radial null geodesics, by virtue of Eqs. (\ref{kt1})
and (\ref{kr1}), satisfy
\begin{equation}
\frac{dt}{dr} = \frac{z \pm e^{ \chi} \Pi}{1 \pm z e^{ \vartheta}
\Pi} \label{eq:de1}
\end{equation}
At this point, we note that a curvature singularity forms at the
origin  $r =  0$, where the  physical quantities like density and
pressure diverges.  The singularity is at least locally naked if
there exist radial null geodesics emerging from the singularity,
and if no such geodesics exist it is a black hole. If the
singularity is naked, then there exists a real and positive value
of $z_{0}$ as a solution to the algebraic equation \cite{psj}
\begin{equation}
z_{0} = \lim_{t\rightarrow 0 \; r\rightarrow 0} z =
\lim_{t\rightarrow 0 \; r\rightarrow 0} \frac{t}{r}=
\lim_{t\rightarrow 0 \; r\rightarrow 0} \frac{dt}{dr}
\label{eq:lm1}
\end{equation}
Using (\ref{eq:de1}) and L'H\^{o}pital's rule we can derive the
following equation
\begin{equation}
V(z_0)\Pi(z_0) = 0 \label{eq:pe}
\end{equation}
Since $\Pi>0$, this implies that
\begin{equation}
V(z_0) = 0 \label{eq:pe3}
\end{equation}
This algebraic equation governs the behavior of the tangent near
the singular points.  The central shell focusing is at-least
locally naked  if Eq. (\ref{eq:pe3}) admits one or more positive roots
(see also Waugh and Lake \cite{wl}, and Ori and Piran \cite{op}).
The values of the roots give the tangents of the escaping
geodesics near the singularity. Thus the visibility of the
singularity depends on the existence of positive roots to Eq. (\ref{eq:pe3}).

Next, we ask the question when this will be realized in terms of
the parameter in self-similar field equations.  This can be
achieved by further analysis of the self-similar field equations.
To this end, we define two new functions $y= z^{\beta}$ and $U^2=
{e^{\chi -  \vartheta}}/{z^2} = y^{-n}\zeta^{-n \beta}S^{-2 n}$.
Here $0\leq \alpha \leq1$, $\delta=1+\alpha$ and $\beta =
{2(1-\alpha)}/{n(1+\alpha)}$

With this transformations Eq.(\ref{sd}) and (\ref{vd}), respectively, become
\begin{eqnarray}
&& \frac{n(n-1)}{2}\left(\frac{S'}{S} \right)^2 \beta^2 y^2
(U^2-1)+ \left(\frac{n}{2}\frac{S'}{S} \right) \beta y
\left[2(n-1)+\frac{2}{n}\delta y^n \zeta^{n \beta/2} U^2 \right]\nonumber \\
&& - \left[\frac{n(n-1)}{2} + \left(1-\frac{n(n-1}{2\zeta
S^2}\right) y^n \zeta^{n \beta/2} U^2 \right]=0 \label{sd1}
\end{eqnarray}
\begin{equation}
\beta y \frac{\zeta'}{\zeta} = \frac{1}{U^2-\alpha} \left( 2\alpha
- n \delta \beta y U^2 \frac{S'}{S}-\frac{1}{n} \delta^2 y^n
\zeta^{n \beta/2} U^2 \right) \label{sd2}
\end{equation}
We analyze the above differential equations near the point $y=
y_0= y(z_0)$ with the condition that $U(y_0)= (\zeta_0 z_0)^{-n
\beta} S_0^{-2 n} = 1$.

Let us write
\begin{equation}
\zeta(z) = \zeta_0 + \zeta_0 \sum_{k=1}^{\infty} \zeta_{k}
(y-y_0)^k \label{es}
\end{equation}

\begin{equation}
S(z) = S_0 + S_0 \sum_{k=1}^{\infty} S_{k} (y-y_0)^k \label{ss}
\end{equation}

which gives $\zeta'(y_0) = \zeta_0 \zeta_1$ and $S'(y_0) = S_0 S_1$. Eqs.
(\ref{sd1}) and (\ref{sd2}) can be recast as

\begin{equation}
\zeta_{1} = \frac{1}{\beta y_0 (1-\alpha)} \left(2\alpha - n \beta
\delta y_0 S_1 - \frac{1}{n} \delta^2 y_0^n \zeta_0^{n \beta/2}
\right) \label{et1}
\end{equation}

\begin{equation}
\beta y_0 S_1 = \frac{n}{n (n-1) + \delta y_0^n \zeta_0^{n
\beta/2}}\left[ \frac{n-1}{2} + \frac{1}{n} \left(1 - \frac{n (n-1)}{2 \zeta_0
S_0^2}\right) y_n \zeta_0^{n \beta/2} \right] \label{ss1}
\end{equation}

Eliminating $S_1$ and $S_0$ from the above equations leads to
\begin{eqnarray}
&& Y^{2 n} + \left[m w-\frac{n^2(n-1)}{2} w \right] Y^{n+1} +
\left[\frac{n(n-1)}{2}\delta - 2 n \alpha +n^2 \right] \nonumber \\
&& \times Y^n+ n (n-1) \delta m w Y + \frac{n^3 (n-1)}{2} \delta^2
- 2 n^2 (n-1) \alpha \delta = 0 \nonumber \\
&& \label{ae}
\end{eqnarray}
where $Y = \delta^{2/n} \zeta_0^{\beta/2} y_0$, $m =
(\beta(1-\alpha)\zeta_1)/{\zeta_0^{\beta-1}}$ and  $ w= {n
\zeta_0^{(\beta-2)/2}}/{\delta^{2/n}}$. This algebraic equation
ultimately decides the final fate of the collapse.  In general,
the existence of real  positive  roots of above algebraic equation
will put a limitation on the physical parameters $\zeta_0$ and
$\zeta_1$.  Thus the existence of  a real positive root of
$V(z)=0$ (and, hence existence of a naked singularity) is
characterized by the values of physical parameters $\zeta_0$ and
$\zeta_1$. It is easy to check that the above equation can admit
at the most four real positive roots for all n. In similar
situation of , one gets a quartic equation \cite{dj}.

A naked singularity can be considered as a physically significant
singularity, if it can escape from singularity to far away
observers for a finite period of time. The analysis  to show this
is similar to that of
 4D case \cite{dj}.  Here we shall restrict
ourselves to main results and readers are urged to refer original
reference for the details of the mathematical description.

The singularities are visible for a finite period of time, if
infinity of integral curves escape from the singularity. To see
this, we write the equation of geodesics in the from $r = r(z)$.
Using Eqs. (\ref{kt1}) and (\ref{kr1}) we obtain,
\begin{equation}
\frac{dz}{dr} =  \frac{V(z) \Pi(z)}{r [1 + z e^{\vartheta} \Pi]}
\label{gv3}
\end{equation}
We have demonstrated that a singularity to be naked, $V(z)=0$ must
have at least one real positive root. Let $z=z_0$ be a simple root
of $V(z) = 0$.  We could then decompose V(z) as
\[
V(z) = \frac{2}{n}(z-z_0) z_0  e^{ \vartheta(z_0)}[\Theta_0 - n ]
\]
Where $\Theta_0 = \Theta(z_0)$. On inserting this expression of
$V(z)$ in Eq. (\ref{gv3}) and integrating, we arrive at
\begin{equation}
r =\Xi (z-z_0)^{n/[\Theta_0 -n]} \label{sr2}
\end{equation}
where $\Xi$ is the integration constant that labels the different
geodesic. Thus, when $\Theta_0 > n$, an infinity of integral
curves with tangent $z=z_0$ would escape from the singularity.  In
corresponding 4D case integral curves meet the singularity in past
if $\Theta_0 > 2$ \cite{dj}. Further, as in 4D case, it can be
shown that an infinity of integral curves will escape from
singularity if $0 < \Theta_0 < \infty $.  It is seen that
$\Theta_0 > 0$ holds only if the weak energy condition is true.
Thus we have shown that an infinity of integral curves would
escape the singularity provided the weak energy condition is
fulfilled.

The naked singularities discussed in previous section are locally
naked, i.e., invisible to an asymptotic observer.  Next, we briefly
describe, when geodesics can reach  an asymptotic observer.  This follows
from the analysis of the Eq. (\ref{gv3}), which can be rearranged
as
\begin{equation}
r = \Xi \; exp \left[ \int \frac{[1+ z e^{\vartheta} \Pi]}{(z-a_1)
\psi(z)} dz - \int \frac{[1+ z e^{\vartheta} \Pi]}{(z-a_2)
\psi(z)} dz \right] \label{sr3}
\end{equation}
where
\[
\psi(z) = \frac{(a_1 - a_2) V(z) \Pi(z)}{(z-a_1) (z-a_2)}
\]
where $a_1$ and $a_2$ (with $a_1 > a_2$) are two consecutive roots
of the equation  $V(z)=0$.  Thus all integral curves can meet the
singularity in past with tangent $z=a_1$, and $r= \infty$ can be
realized in future along same integral curves at $z= a_2$. Thus
the singularities will be globally naked and an infinity of curves
would emanate from the singularity to reach a distant observer.

 Finally, we need to
determine the curvature strength of naked singularities, which is
an important aspect \cite{ft1}.  A singularity is gravitationally
strong or simply strong in the sense of Tipler \cite{ft} if every
volume element is crushed to zero dimensions at the singularity,
and weak otherwise (i.e., if it remains finite). It is widely
believed
 that a space-time does not admit an extension through a singularity if it is
a strong curvature singularity in the sense of Tipler \cite{ft}. A
necessary and sufficient measure for a singularity to be strong
has been given by Clarke and Kr\'{o}lak \cite{ck} that for at
least one non-spacelike geodesic with affine parameter $k$, in the
limiting approach to the singularity, we must have
\begin{equation}
\lim_{k\rightarrow 0}k^2 \psi = \lim_{k\rightarrow 0}k^2 R_{ab}
K^{a}K^{b} > 0 \label{eq:sc}
\end{equation}
where $R_{ab}$ is the Ricci tensor.  Our purpose here is to
investigate the above condition along future directed radial null
geodesics that emanate from the naked singularity. Eq.
(\ref{eq:sc}) can be expressed as
\begin{equation}
\lim_{k\rightarrow 0}k^2 \psi = \lim_{k\rightarrow 0} k^2
\frac{(\zeta+p)C^2 e^{\vartheta}[z + e^{\chi} \Pi]^2 }{r^4 [e^{
\chi} - z^2 e^{\vartheta}]^2}  \label{eq:sc1}
\end{equation}
Using Eqs. (\ref{kt1}), (\ref{kr1}), and L'H\^{o}pital's rule, Eq.
(\ref{eq:sc1}) turns out to be
\begin{equation}
\lim_{k\rightarrow 0}k^2 \psi = \frac{n^2
\Theta_0}{(\Theta_0+n)^2}
> 0 \label{sc2}
\end{equation}
Thus, along radial null geodesics strong curvature condition is
satisfied if $\Theta_0 > 0$, which is also a necessary condition
for the energy condition. Thus it follows that singularities are
gravitationally strong if the weak energy condition is satisfied.

\section{Discussion and concluding remarks}
General relativity was formulated in space-time with four
dimensions, of course.  However, there are theoretical hints that
we might live in a world with more than . A generalization of
general relativity to HD has been of considerable interest in
recent times. For further progress toward an understanding of
spherical collapse, from the viewpoint of the cosmic censorship,
one would like to know the effect of extra dimensions on the
existence of a naked singularity.  The relevant questions would
be, for instance, whether such solutions remain naked with the
introduction of extra dimensions. Do they always become covered?
Does the nature of the singularity change? Our analysis shows that
none of the above hold. Infact, the gravitational collapse of
perfect fluid in HD self-similar spherically symmetric space-times
lead to strong curvature globally naked singularities.

It is shown that the singularity at $t=0$, $r=0$ is atleast
locally naked when $V(z_0) = 0$ has a simple positive root and a
infinity of future directed
 integral curves escape from the singularity provided $0 < \Theta_0 < \infty$.
Further such a singularity could be globally naked as well
provided $V(z_0) = 0$ has atleast two simple positive roots.
Further along null rays emanating from the singularity, the strong
curvature condition (\ref{eq:sc}) is satisfied.  Thus the naked
singularities here could be considered as a physically significant
singularity, and hence cannot be ignored. Thus, this offers a
serious counter-evidence to the conjecture.

As mentioned earlier, the conjecture has yet no proof for either
version. However, It appears to be a growing body of evidence
against the conjecture. These examples showing the occurrence of
naked singularities remain important and may be valuable if one
attempts to formulate the notion of the conjecture in precise
mathematical form. They may add some insight  in issues involved
in cosmic censorship. Indeed, the  singularity arising in massless
scalar field led to a formulation of weak censorship \cite{cd}.
Thus there may be much to be learned from studying such counter
examples.

To sum up, this generalizes the previous studies of self-similar
spherical gravitational collapse in  to $D = n+2$ dimensional
space-times and when $n=2$ one recovers the results of analogous
study in 4D.

\section*{Acknowledgments}
Authors would like to thank IUCAA, Pune for hospitality while this
work was done.


\begin{thebibliography}{0}
\bibitem{rp} R. Penrose,  Riv.  Nuovo Cimento {\bf 1}, 252
 (1969); in {\it General Relativity, an Einstein Centenary
 Volume}, edited by S. W. Hawking and W. Israel (Cambridge
 University Press, Cambridge, England, 1979).
 \bibitem{psj} P. S. Joshi, {\it Global Aspects in Gravitation and
Cosmology} (Clendron, Oxford, 1993); P. S. Joshi, {Pramana} {\bf
55}, 529 (2000).
\bibitem{r2} C. J. S. Clarke, { Class.
Quantum Grav.} {\bf 10}, 1375 (1993);  R. M. Wald, gr-qc/9710068;
S. Jhingan and G. Magli, gr-qc/9903103;   T. P. Singh, {J.
Astrophys. Astron.} {\bf 20}, 221 (1999).
\bibitem{op}  A. Ori and T. Piran {Phys. Rev. Lett.} {\bf 59},
2137 (1987).
\bibitem{op1}   A. Ori and T. Piran {Phys. Rev. D} {\bf 42},
1068 (1990).
\bibitem{wl} B. Waugh and K. Lake, {Phys. Rev.} {\bf D 38},
1315 (1988); {Phys. Rev.} {\bf D 40}, 2137 (1989); K.Lake and T.
Zannias {Phys. Rev. D} {\bf 41}, 3866 (1990).
\bibitem{hp} R. N. Henriksen and K. Patel, {Gen. Rel. Grav.}
 {\bf 23}, 527 (1991).
 \bibitem{dj}P. S. Joshi and I. H. Dwivedi, {Commun. Math. Phys.}
 {\bf 146}, 333 (1992); Lett. Math Phys. {\bf 27}, 235 (1993).
\bibitem{jl} J. P. S. Lemos, {Phys. Lett. A}
{\bf 158}, 271 (1991).
\bibitem{bn} B. C. Nolan {Class. Quantum Grav. }
{\bf 18},  1651 (2001).
\bibitem{rs} L. Randall and R. Sundram, {Phys. Rev. Lett.}
{\bf 83}, 3370 (1999); {\bf 83}, 4690 (1999).
\bibitem{hd} A. Banerjee, A. Sil and S. Chatterjee {Astrophys. J.}
 {\bf 422}, 681 (1994); A. Sil and S. Chatterjee {Gen. Relativ. Gravit.}
 {\bf 26}, 999 (1994); J. Soda and K. Hirata, {Phys. Lett. B}
 {\bf 387}, 271 (1996); A. Ilha and J. P. S. Lemos,  {Phys. Rev. D}
 {\bf 55}, 1788 (1997); A. Ilha, A. Kleber and J. P. S. Lemos, {J. Math. Phys.}
{\bf 40}, 3509 (1999);  A. V. Frolov {Class. Quantum Grav.} {\bf
16}, 407 (1999);  J. F. V. Rocha and  A. Wang  {Class. Quantum
Grav.} {\bf 17}, 2589 (2000).
\bibitem{gd} S. G. Ghosh and N. Dadhich, {Phys. Rev. D} {\bf 64},
047501 (2001).
\bibitem{gb} S. G. Ghosh and A. Beesham, {Phys. Rev. D} {\bf 64},
124005 (2001).
\bibitem{gss} S. G. Ghosh,  S. B.~Sarwe and R. V.~Sararykar, {accepted in Phys. Rev. D}
\bibitem{ct} M. E. Cahill and A. H. Taub {Commun. Math. Phys.} {\bf 21},
1 (1971).
\bibitem{he} S. W. Hawking and G. F. R. Ellis, {\it The Large Scale Structure
of Space-time} (Cambridge University Press, Cambridge, 1973).

\bibitem{no} N. O.~Santos {Mon. Not. R. Astr. Soc.} {\bf 216},
403 (1985).
\bibitem{rw} J. F. V. Rocha, A. Wang and N. O.~Santos
  {Phys. Lett. A} {\bf 255}, 213 (1999).

\bibitem{cm} M.E.Cahill and G. C. McVittie  {J. Math. Phys.} {\bf
11}, 1382 (1970)
\bibitem{ft1} F. J. Tipler {Phys. Lett. A} {\bf 64},
8 (1987).
\bibitem{ft} F. J. Tipler, C. J. S. Clarke, and G. F. R. Ellis in
{\it General Relativity and Gravitation}, edited by A Held
(Plenum, New York, 1980).
\bibitem{ck} C. J. S. Clarke and  A. Kr\'{o}lak {J. Geom. Phys.} {\bf
2}, 127 (1986).
\bibitem{cd} D. Christodoulou { Ann. Math.} {\bf 149},
183 (1999).
\end{thebibliography}
\end{document}